\documentstyle[aps,preprint]{revtex}
\begin{document}
\draft
\title{Avalanches in the Weakly Driven Frenkel-Kontorova Model}
\author{Franz-Josef Elmer}
\address{Institut f\"ur Physik, Universit\"at
   Basel, CH-4056 Basel, Switzerland}
\maketitle
\begin{abstract}
A damped chain of particles with harmonic nearest-neighbor
interactions in a spatially periodic, piecewise harmonic potential
(Frenkel-Kontorova model) is studied numerically. One end of the chain is
pulled slowly which acts as a weak driving mechanism. The numerical
study was performed in the limit of infinitely weak driving. The model
exhibits avalanches starting at the pulled end of the chain.
The dynamics of the avalanches and their size
and strength distributions are studied in detail.
The behavior depends on the value of the damping constant.
For moderate values a erratic sequence of avalanches of all sizes occurs.
The avalanche distributions are power-laws which is a key feature of
self-organized criticality (SOC). It will be shown that the system
selects a state where perturbations are just able to propagate
through the whole system. For strong damping a regular
behavior occurs where a sequence of states reappears periodically
but shifted by an integer multiple of the period of the
external potential.
There is a broad transition regime between regular and irregular behavior,
which is characterized by multistability between regular
and irregular behavior.
The avalanches are build up by sound waves and shock waves.
Shock waves can turn their direction of propagation,
or they can split into two pulses propagating in opposite
directions leading to transient spatio-temporal chaos.
\end{abstract}
\pacs{PACS numbers: 05.70.Ln,05.50.+q,46.10.+z%
\\Appears in Phys. Rev. E.%
}

\narrowtext
\section{Introduction}\label{s1}

The Frenkel-Kontorova (FK) model \cite{r1} (i.e. a harmonic chain in
an external spatially periodic potential) is a favorite model for
disordered systems and glassy materials because of its {\em huge
number of meta-stable states\/}. The following topics are intensively
investigated in the literature: (i) the commensurate-incommensurate
transition of the ground state \cite{r2}, (ii) the lowest excitations
in order to calculate the specific heat \cite{r3}, (iii) the
over-damped dynamics for a strongly tilted external potential in
order to understand the pinning-depinning transition of
charge-density waves \cite{r4}.

In 1987 Bak, Tang, and Wiesenfeld \cite{r5} showed that in {\em
weakly driven} dissipative systems a huge number of meta-stable
states can lead to a new kind of erratic behavior, called {\em
self-organized criticality} (SOC). On the phenomenological level the
key feature of SOC is an erratic sequence of events (called {\em
avalanches}) on all scales, i.e. a power-law distribution of the
strength of the event. Usually this phenomenon is accompanied by a transient
where the distribution does not necessarily show a power-law. Finally the
system reaches the so-called {\em SOC attractor\/}. Bak et al. found
this behavior in a cellular automaton (dimension $\ge 2$)
which mimics a sand pile. Later,
this and other cellular automata as well as coupled maps were
studied in order to investigate SOC \cite{r6}. In order to explain
the observed power-law distributions Bak et al. proposed the following
mechanism: the weak driving moves the system to the
``edge'' of the stability region where it is just critical and
therefore exhibits no characteristic scales. The edge represents a
threshold for the propagation of a perturbation through the whole system.
The crucial point that Bak et al. assumed is that a state outside the
stability region will stick just at the stability edge during its
relaxation towards the stability region. It is interesting to note
that in their seminal paper \cite{r5} Bak et al. used the {\em
over-damped\/} FK model in order to demonstrate these properties.
But they rejected this system for further studies because
always the same so-called ``least-stable'' state reappears, a feature
which is shared by their one-dimensional cellular automaton in
contrast to the higher dimensional cellular automata that display SOC
behavior. Nevertheless, SOC is
possible in the FK model if {\em inertia} is taken into account
\cite{r7}.

The present paper presents results of a detailed study of the weakly
driven FK model. The driving mechanism is a slow pulling of the chain
at one end.  Our study sheds some new light on the complex behavior
of the FK model because it is treated in a regime far away from
previous investigations: (i)~the driving force pushes the system into
meta-stable states with large energies; (ii)~this driving mechanism
does not decrease the number of meta-stable states as it is usually
the case in depinning-pinning transition studies; (iii)~the dynamics
is neither conservative nor over-damped.

To be more specific, a chain of $N+1$ particles, $i=0,1,\ldots,N$,
of mass $m=1$ with nearest-neighbor interaction $V_I$ in a spatially
periodic external potential $V_E$ is investigated. We assume that all
particles are damped with the same damping constant $g$. The equation of
motion is therefore:
\begin{eqnarray}
  \ddot{x_i}+g\dot{x_i}&=&V_I'(x_{i-1}-x_i)-V_I'(x_i-x_{i+1})\nonumber\\
   &&-V_E'(x_i),\quad i=1,\ldots,N.
   \label{eqm}
\end{eqnarray}
Particle $i=0$ is pulled (or pushed) with constant velocity
\begin{equation}
  x_0 = v t\quad\mbox{with}\quad |v|\ll 1,
  \label{x0}
\end{equation}
and particle $i=N$ is the free end of the chain. The latter condition
is taken into account by introducing an $(N+2)$nd variable $x_{N+1}$
chosen in such a way that $V'_I(x_N-x_{N+1})=0$. Recently
simulations of the original FK model \cite{r7} (i.e.
$V_I(x)=(x-a)^2/2$, $V_E(x)=b\cos x$) have been performed, as well as
of a FK model with Toda-like nearest-neighbor interaction \cite{r8} (i.e.
$V_I(x)=e^{-x}+e^{-a}x$, $V_E(x)=b\cos x$) serving as a simplified
model of a ferromagnetic Bloch-wall array in a spatially periodic
field. In this paper the FK model with a piecewise parabolic
external potential (see Fig.~\ref{f1}) is investigated, i.e.
\begin{eqnarray}
  V_I(x)&=&(x-a)^2/2 \label{vi}\\
  V_E(x)&=&b\,(\case{1}{2}-x\,\mbox{mod}\,1)^2,\quad b>0. \label{ve}
\end{eqnarray}

The general behavior in all cases is the following: the slow
pulling (or pushing in the case of the Toda potential) locally puts
energy into the system at a very low rate. This drives the system
towards an instability at which at least a fraction of this energy is
released and propagates along the chain. Since the chain is
damped, the energy dissipates and the system settles down in one of
the huge number of meta-stable states. We will call such an event an
{\em avalanche}. It rearranges the configuration of the chain at
least locally. Previous studies \cite{r7,r8} indicate the existence
of power-laws
for the distributions of avalanche strengths measured by taking the
sum over all particle displacements during the avalanche. In the
overdamped limit the sequence of avalanches becomes regular as
expected by Bak et al. \cite{r5}.

The external potential in the FK model can be interpreted as a
regular array of pinning centers of equal strength for some objects
(e.g. charge-density waves, flux lines, or Bloch walls).
SOC-like behaviour also has been shown to occur in overdamped models with
randomly distributed pinning centers \cite{r20,r21}. In \cite{r20} the same
driving mechanism as in our model was used (i.e. pulling the chain at
one end). The erratic and SOC-like behavior in those models seems
to be caused by the randomness of the external potential. In our case
the inertia together with the nonlinear character of the external
potential cause irregular avalanches (see Sec.~\ref{s5}) which seems
to be responsible for the same type of behavior.

To compare the results with the behavior of cellular automaton models
showing SOC it is important to drive the system very slowly because
deviations are expected from finite driving. In other words, a
separation of times scales is necessary: a fast time scale is given
by the averaged duration of an avalanche, and the slow time scale is
defined as the averaged time interval between the occurrence of successive
avalanches.
Infinitesimally slow driving means that the slow time scale tends to
infinity because on average two avalanches are separated by an
infinite time interval.
The limit of infinitesimally slow driving cannot be reached
in computer simulations of continuous systems
described by differential equations. Therefore a finite but very
small driving is usually chosen to get a compromise between time scale
separation and computer time consumption.

The main advantage of a piecewise parabolic external potential is the
possibility to do the simulations in the limit of {\em infinitesimally\/}
slow drive, i.e. $v\rightarrow 0$. This is due to the fact that all
stationary states are uniquely characterized by the integer part $n_i$ of the
particle positions $x_i$ which we will call {\em well number\/}. In the
next section we show that it is possible to reconstruct the stationary
states from the position of the fixed end $x_0$ and the well numbers.
Furthermore, we are able to calculate the stability interval of
$x_0$, i.e. the interval in which an adiabatic move of $x_0$ does
not lead to an avalanche. Therefore, in a computer simulation we
abridge the adiabatic move by putting the system into the state at
the edge of the next instability.

There is another reason for choosing a piecewise parabolic potential:
it is well known from the work of Aubry \cite{r2} that in a smooth
potential with potential strength below the so-called ``analyticity
breaking point'' the ground state can be shifted without energy
because a Goldstone mode exists. Therefore we would expect that no
avalanches occur and the chain would creep over the
``washboard'' $V_E$. It is also well-known that the loss of the
Goldstone mode is due to the disappearance of the so-called ``last
KAM trajectory'' in the two-dimensional symplectic map which
describes all stationary states. The piecewise parabolic
potential leads to a purely hyperbolic map which do not have KAM
trajectories.

A model similar to the FK model is the Burridge-Knopoff (BK) model
\cite{r12}, which was proposed as a model of an earthquake fault. In
the last years various modifications of it have become popular in
order to investigate SOC \cite{r9a,r9b,r10,r11}. The main difference to
the FK model is that the external potential and damping are replaced
by phenomenological stick-slip friction forces. A particle is at rest
if the forces resulting from the springs are weaker than the static
friction force. Originally the chain is driven by weak springs
connecting each particle with a rigid and slowly moving plate. A
driving mechanism similar to that one in our model was also studied
(train model) \cite{r11}. The external potential together with the
damping in the FK model can be considered as a microscopic model for
a stick-slip friction law. It has the properties \cite{r17} that
(i) the dynamic friction increases with velocity,
(ii) the static friction is larger than the dynamic friction if
the damping constant $g$ is not too large. In that case the friction force
is therefore discontinuous at zero. Avalanches in the BK model are
possible only if the dynamic friction is less than the static
friction. Otherwise only creeping is
possible \cite{r22}. Other details (e.g. whether the dynamic friction
decreases or increases with velocity) of the various phenomenological
friction laws are unimportant.

The paper is organized as follows: Sect.~\ref{s2} deals with
stationary states. It gives the tools for solving the problem of
infinitesimally slow driving in the simulations. Details of
the whole simulation are given in Sec.~\ref{s2a}. Sec.~\ref{s4}
gives a detailed analysis of the SOC attractor. Sec.~\ref{s3}
presents avalanche statistics and attractor dynamics for various
values of the model parameters. Sec.~\ref{s6} presents details of
the avalanche dynamics, and also discusses the mechanism which leads to
SOC in the FK model. Sec.~\ref{s5} investigates the transition to
the over-damped case. The concluding section compares the
results obtained here with results from other models.

\section{Stationary states}\label{s2}

This section deals with the stationary states of the FK model with
piecewise harmonic external potential. The methods for (i) generating
stationary states, (ii) obtaining compact descriptions (``symbolic
dynamics'') of them, and for (iii) calculating points of instability
are developed. Furthermore the number of meta-stable states is
calculated.

The stationary states for a prescribed value of $x_0$ are given by
the solutions of
\begin{eqnarray}
  0&=&x_{i+1}-2x_i+x_{i-1}\nonumber\\
  && -b[2(x_i\,\mbox{mod}\,1)-1],\quad i=1,\ldots,N,
  \label{eqss}
\end{eqnarray}
where $x_{N+1}$ is defined by $x_{N+1}\equiv x_N-a$. It should be noted
that every solution of (\ref{eqss}) is either stable or meta-stable,
i.e. no linearly unstable solutions exist. This can be shown by
calculating the eigenvalues of the matrix given by the second variation
of the potential energy. The eigenvalues (i.e. the phonon dispersion
relation)
\begin{equation}
  \omega_k^2=2(1+b-\cos k)
  \label{wk}
\end{equation}
are always positive. Therefore the matrix is positive definite which
proves the stability.

A very general way to solve Eq.~(\ref{eqss}) is the following (it is
also applicable to other FK models\cite{r7,r8,r20}): choose an
arbitrary value of $x_N$ and then calculate iteratively the sequence
$x_{N-1},x_{N-2},\ldots,x_1,x_0$. Change $x_N$ slightly until
$x_0$ matches its prescribed value. This is not an effective way to
calculate {\em all} stationary states but it shows that {\em each}
stationary state is {\em uniquely} determined by $x_N$. Furthermore,
we can think of $x_0$ as a function of $x_N$. For
$N\rightarrow\infty$ this function will be nowhere differentiable and
the cut of it with a horizontal line gives a Cantor-like set. Every point
in this set corresponds to a certain stationary state.

In our case $x_0(x_N)$ is piecewise linear since the external
potential is piecewise harmonic. The pieces are produced by the modulo
term $x_i\,\mbox{mod}\,1$ of Eq.~(\ref{eqss}) at each step of the iterative
computation of this function. They all have the slope $(1+2b)^N$ and
they are separated by jumps of height $2b$ (see Fig.~\ref{f2})
\cite{foo1}. The lower and upper value of each piece determines the
interval inside which $x_0$ can vary slowly without releasing an
avalanche. At the boundary of that interval some particle sits
exactly on a cusp of the external potential. This particle will be
destabilized if $x_0$ is moved slightly beyond the boundary.  The
important point is that we are able to calculate this stability
interval for a given stationary state.

The key for an effective method to generate all stationary states for
a prescribed $x_0$ is the existence of a compact description, called
``symbolic dynamics'', for all stationary states. Each description is
a list of exactly $N$ symbols which determines {\em uniquely\/} a
stationary state. We present two of them.

In the first symbolic dynamics the symbols, called {\em well numbers\/}
(see Fig.~\ref{f1}), are the integer parts $n_i\equiv
\mbox{Int}(x_i)$ of the positions $x_i$ \cite{r13}. By applying the
Greens function method on $2(b+1)x_i-x_{i+1}-x_{i-1}=b(1+2n_i)$ it is
easy to show that the positions $x_i$ can be reconstructed uniquely
by $x_0$ and the well numbers $n_i$:
\begin{equation}
 x_i=C\sum_{j=1}^N(1+2n_j)\eta^{|i-j|}+\alpha_0\eta^i+\alpha_N\eta^{N-i},
  \label{ss}
\end{equation}
with
\begin{equation}
  C=\frac{b/2}{\sqrt{b(b+2)}},
  \label{C}
\end{equation}
\begin{equation}
  \eta=1+b-\sqrt{b(b+2)}.
  \label{eta}
\end{equation}
The coefficients $\alpha_0$ and $\alpha_N$ are given by the solutions of
\begin{mathletters}
\label{alpha}
\begin{eqnarray}
   \alpha_0+\eta^N\alpha_N&=&x_0-C\sum_{i=1}^N(1+2n_i)\eta^i,\\
   -\eta^N\alpha_0+\eta^{-1}\alpha_N&=&-\frac{a}{1-\eta}+
      C\sum_{i=1}^N(1+2n_i)\eta^{N-i}.
\end{eqnarray}
\end{mathletters}
Clearly Eq.~(\ref{ss}) does not hold for arbitrary well numbers. Only
such well number configurations are allowed which fulfill the
self-consistency condition
\begin{equation}
  n_i = \mbox{Int}(x_i),\quad i=1,\ldots,N.
  \label{selfcon}
\end{equation}
If a stationary state is given we obtain a compact description of it by
calculating the well numbers. Using (\ref{ss}) we can completely reconstruct
the stationary state from the well numbers, i.e. there is a one-to-one
relation between $\{x_1,\ldots,x_N\}$ and $\{n_1,\ldots,n_N\}$.

There is an even more compact description by using the so called
{\em $f$ symbols\/} (see Fig.~\ref{f1}) defined by
\begin{equation}
  f_i = n_{i+1}-2n_i+n_{i-1},\quad i=1,\ldots,N,
  \label{ntol}
\end{equation}
with $n_{N+1}\equiv n_{N}-\mbox{Int}(a)$ and $n_0=\mbox{Int}(x_0)$.
The $f$ symbols give a rough estimate of the resultant force on the
particles due to the springs (i.e. $x_{i-1}-2x_i+x_{i+1}$ which should
be less than $b$). Using (\ref{eqss}) we obtain the following inequality
\begin{equation}
  |f_i| \leq \mbox{Int}(2+b).
  \label{ineql}
\end{equation}
For given $f$ symbols we get for the well numbers
\begin{equation}
  n_i=\mbox{Int}(x_0)-i\,\mbox{Int}(a)-\sum_{j=1}^N\min(i,j)\,f_j.
  \label{lton}
\end{equation}
For particles in the bulk of the chain the position relative to the
potential well can be calculated directly from the $f$ symbols
\begin{equation}
x_i\,\mbox{mod}\,1=x_i-n_i=\frac{1}{2}+\frac{C}{b}\sum_jf_{i+j}\eta^{|j|}.
  \label{ssl}
\end{equation}
This formula is obtained from Eqs.~(\ref{ss}) and (\ref{lton}) by
shifting the limits of the sums to $\pm\infty$ and by dropping the
$\alpha$-terms in (\ref{ss}).

Next we describe a method for generating an arbitrary stationary
state. As a byproduct we also get the stability interval. The method
simply iterates intervals by using Eq.~(\ref{eqss}). We start with
the interval $[0,1)$ assuming well number zero for the $N$-th
particle. One iteration step leads to the interval $[-a-b,1-a+b)$
which gives the possible positions of particle number $N-1$. In the
next iteration the modulo term in Eq.~(\ref{eqss}) cuts this interval
into subintervals since its length is greater than unity. Because each
subinterval should be treated separately we have to choose which
interval we want to iterate further. The subintervals are uniquely
labeled by the well number because the cuts occur at integer points.
This iteration scheme leads immediately to the stability interval
$(x_0^{left},x_0^{right})$ because after we have chosen the well
number of particle $i=1$ we get the interval of all possible positions
of particle $i=0$ (i.e. the fixed particle) after the next iteration
step.

We finish this section with the calculation of the numbers of
meta-stable states. With the method of the last paragraph the total
number of meta-stable states $M(N)$ can be easily calculated numerically.
For large $N$ this number increases exponentially with $N$.
The rate of increase is the {\em topological entropy} $\nu$, i.e.
\begin{equation}
  \nu = \lim_{N\rightarrow\infty}\frac{\ln M(N)}{N}.
  \label{te}
\end{equation}
We can calculate $\nu$ {\em analytically} with the following
consideration: first we rewrite Eq.~(\ref{eqss}) in the form
\begin{equation}
  \begin{array}{rcl}
  d_{i+1} &=& d_i + b[2(x_i\,\mbox{mod}\,1)-1]\\
  x_{i+1} &=& x_i + d_{i+1}.
  \end{array}
  \label{map1}
\end{equation}
This is a two-dimensional symplectic map similar to
Arnold's cat map \cite{r14} ($b=1/2$ corresponds to the original cat
map). Next we use a theorem of Pesin which says that the metric (or
Kolmogorov-Sinai) entropy is given by the sum of the positive
Lyapunov exponents averaged over the invariant density \cite{r14}.
The Lyapunov exponents of the map (\ref{map1}) are $\pm\ln\eta$.
Because the invariant density of the map (\ref{map1}) is constant
\cite{r14} the topological entropy is equal to the metric entropy,
therefore we get
\begin{equation}
  \nu = -\ln\eta = \ln[1+b+\sqrt{b(b+2)}].
  \label{nu}
\end{equation}
This result is verified by direct numerical calculations.

\section{Notes about the simulation}\label{s2a}

In order to simulate infinitesimally slow driving (i.e. pulling speed
$v\to 0$) the simulation scheme is broken up into two steps which
are repeated as often as desired.
\begin{description}
\item[Driving step:]
For a given stationary state first the well numbers $n_i$ and the
stability interval $(x_0^{left},x_0^{right})$ of the position of the
fixed end of the chain $x_0$ is calculated accordingly to Sec.~\ref{s2}.
Then $x_0$ is put near the edge but still within the stability interval,
i.e. $x_0=x_0^{right}-\epsilon$, where $\epsilon\ll 1$.
Without changing the well numbers the new positions
$x_i$ are calculated according to Sec.~\ref{s2}. Only the
positions of the particle near the fixed end are noticeably shifted
because the change is proportional to $\eta^i$. Finally, $x_0$ is put
just outside the stability interval, i.e. $x_0=x_0^{right}+\epsilon$.
In all simulations we have chosen $\epsilon=10^{-4}$.
\item[Relaxing step:]
For the simulation of the avalanche dynamics we use a
predictor-corrector scheme described in Ref. \cite{r15} which is
superior to the well-known Verlet algorithm \cite{r16} especially for
potentials with cusp-like singularities \cite{r15}. For the time step
$\Delta t$ of the integration scheme we have chosen 0.01, 0.005, and
0.001 for $b=1$, $b=5$, and $b=20$, respectively. These time steps are
smaller at least by a factor 200 than the fastest oscillation periods
given by the maximum of the phonon dispersion relation (\ref{wk}).

Because of dissipation of energy the avalanche dies out. We use the
following criterion for the death of an avalanche: the barrier for a flip
of a single particle into a neighboring potential well is calculated
under the assumption that the neighboring particles are fixed.
Because this barrier height is only approximately correct we require
that the kinetic energy should be less than one tenth of this barrier
height. If this criterion is fulfilled for all particles the simulation
for a single avalanche is stopped.
\end{description}

There is a strong similarity to cellular automata showing SOC \cite{r6}
where also a distinction is made between driving rules and relaxing
rules which are applied to stable states and unstable states,
respectively. These automata can be seen to be driven in a
``infinitely slow'' limit because the driving rules are applied not
before even the largest avalanche has died out.

We end this section with some remarks about the values we have chosen
for the four system parameters $a$, $b$, $N$, and $g$. In the study
of the ground state of the FK model the ratio between equilibrium
length of the springs and potential periodicity plays an important
role \cite{r2}.  In our case this ratio is unimportant, because the
chain is driven far away from the ground state (see next section).
Only the strength and the probability of the smallest avalanches are
slightly changed.  In fact most of the energy is stored in the
springs.  We have chosen the following values: equilibrium length
$a=1.3$; strength of the external potential $b=1$, 5, and 20; number
of particles usually $N=125$ and 500; damping constant
$g=10^{-3}\ldots 10$.  If not mentioned otherwise, all results are
obtained from simulation runs of $2\cdot 10^4$ avalanches except for
$b=20$ where the number of avalanches was only $10^4$.

\section{The SOC attractor}\label{s4}

The separation of time scales caused by the infinitesimally slow
driving leads to the definition of {\em two\/} phase spaces: the
first one is the usual phase space of a mechanical system defined by
the positions and momenta of all particles (including the zeroth
particle). The second phase space called {\em reduced phase space\/}
--- a sub-space of the former one --- is defined by all stationary
states modulo an integer amount of the potential periodicity. A point
of it is given by the $f$ symbols of the stationary state. From
Sect.~\ref{s2} we know that the particle positions can be
reconstructed from the $f$ symbols except that of particle number
zero (the fixed end). This is not a restriction because only the
stability interval $(x_0^{left},x_0^{right})$ is needed which can be
calculated from the $f$ symbols. The reduced phase space is discrete
and finite, similar to the phase space of cellular automata showing
SOC \cite{r6}. Unstable states of these automata  correspond to
unstable states of the reduced phase space (i.e. sequences of $f$
symbols which do not correspond to stationary states of the chain).
The dynamics of an avalanche takes place in the full phase space and
cannot be uniquely projected onto the reduced phase space.

Similar to the distinction between two phase spaces we distinguish
between the usual {\em physical time\/} $t$ and an integer {\em
pseudo time\/} $\tau$ which will be increased by unity after each
occurrence of an avalanche. We also distinguish between the {\em full
dynamics\/} taking place in the full phase space and the {\em reduced
dynamics\/} which is a discrete dynamics on the reduced phase space.
The reduced phase space is defined in such a way that an avalanche
shifting the chain by one period of the potential leads to an orbit
of period one. The reduced dynamics can be interpreted as a cellular
automaton with complicated nonlocal rules. From the simulation scheme
described in section~\ref{s2a} we see that the automaton is
completely deterministic. The automaton would be a stochastic one if
we would take into account the unavoidable thermal noise which will be
amplified by the chaotic nature of the avalanche dynamics (see
section~\ref{s6}).

The physical time $t$ is not useful to describe the whole dynamics
since the mean time between successive avalanches goes to infinity
for pulling velocity $v\to 0$. Instead, we use $x_0=v\,t$.  Remember
that during a single avalanche $x_0$ is constant. In Sec.~\ref{s3} we
will see that, on average, $x_0(\tau+1)-x_0(\tau)$ remains finite
even for $v\to 0$. Presumably, it will be independent of $v$ for at
least very small values. Thus knowing the distribution of
$x_0(\tau+1)-x_0(\tau)$ we are able to calculate the distribution of
waiting times between successive avalanches.

Starting with the ground state the driving mechanism will stretch the
chain. Most of the avalanches release less energy than the work due
to stretching which is put into the system between successive
avalanches. This is still true after a transition time where the
system reaches an attractor in the reduced phase space (see
Fig.~\ref{f3}).  The transition time usually ends after the
occurrence of the first avalanche involving all particles of the
chain.

We call this attractor the {\em SOC attractor\/} though the avalanche
distribution may strongly deviate from an expected power law behavior
(see section~\ref{s3}). It is a {\em self-organized\/} balance
between driving which puts energy into the system and avalanching
which dissipates energy. The motion on the SOC attractor looks
erratic although we know that it can only be a periodic orbit because
of the finiteness of the reduced phase space. But the recurrence time
is very large, presumably of the same order as the number of stationary
states. Thus it increases exponentially with the number of particles.

In order to characterize the SOC attractor we first discuss the
distribution of the potential energy
\begin{equation}
  E_{pot}\equiv\sum_{i=1}^NV_I(x_{i-1}-x_i)+V_E(x_i).
  \label{epot}
\end{equation}
The potential energy is calculated at the instability point
$x_0=x_0^{right}$ just before the avalanche starts. The simulations
show that the mean potential energy $\overline{E}_{pot}$ clearly
scales with $N^3$. This scaling behavior can be easily understood if
we assume that, on average, the {\em local stress\/} of the chain
$x_{i-1}+x_{i+1}-2x_i$ is the same for each particle in
the bulk of the chain. Therefore the length of the springs decreases
linearly with the particle index from strongly stretched at the fixed
end to almost equilibrium length at the free end. Using
$\sum_{i=1}^Ni^2=N(N+1)(2N+1)/6$ we get
\begin{equation}
  \overline{E}_{pot}/N^3 = \frac{\overline{f}^2}{6}+{\cal O}(N^{-1}),
  \label{en3}
\end{equation}
where $\overline{f}\equiv\overline{x_{i-1}+x_{i+1}-2x_i}$ is the
mean local stress which is equivalent to the mean $f$ symbol per
particle. Using eq.~(\ref{eqss}) it can be expressed in terms of the
mean position of the particle related to the potential well:
\begin{equation}
  \overline{f}=b\,(2\,\overline{x_i\,\mbox{mod}\,1}-1).
  \label{lx}
\end{equation}
Fig.~\ref{f4} shows that $\overline{x_i\,\mbox{mod}\,1}$ (and
therefore also $\overline{f}$) is independent of $i$.
The deviation between $\overline{f}$ calculated from
Eq.~(\ref{en3}) and from Eq.~(\ref{lx}) is less than 1\%.
Since the states are analysed at the stability boundary $x_0^{right}$
the value 1 for $\overline{x_i\,\mbox{mod}\,1}$ at $i=1$ expresses the
fact that almost all avalanches start at the first particle.

The mean SOC attractor is similar to the attractor obtained by
pulling the chain with finite velocity $v$ through a viscous medium
with no external potential (i.e. $b=0$). Here all particles travel
with $v$. From the equation of motion (\ref{eqm}) we get
$x_{i-1}+x_{i+1}-2x_i=g\,v$. Thus the potential energy scales also
like $N^3$.

Fig.~\ref{f5} shows $\overline{E}_{pot}/N^3$ for several values of
$b$ and $g$. The mean potential energy increases with $b$ and $g$.
The increase with $b$ does not originate from the increase of energy
stored in the external potential, but from the fact that the
stretching of the springs can be increased because of stronger
particle pinning. The approximate power-law in $g$ (exponent $\approx
1.5$) is similar to the power-law behavior of residual energy as a
function of cooling rate expected in glassy materials \cite{r18}.
This expectation was recently confirmed \cite{r19} by simulations of
a mechanical model similar to the FK model (a chain of particles with
anharmonic on-site potential and harmonic nearest-neighbor
interactions). The cooling was simulated by viscous damping (cooling
rate proportional to damping constant); the energy of the initial
state was large enough to allow the system to move in phase space
{\em unhindered\/} by potential barriers. Similar annealing
simulations with the FK model lead to residual energies much below
the mean energy of the SOC attractor because the residual energy
scales only with $N$. On the other hand, simulations starting with
{\em stable\/} initial states with potential energies much larger than the
mean energy of the SOC attractor show that, on average, the energy of
the state after the first avalanche is almost identical with the mean
energy of the SOC attractor (see dotted energy distributions in
Fig.~\ref{f3}) as expected for SOC \cite{r6}.

The distribution of the potential energy of the SOC attractor is
strongly non-gaussian. Fig.~\ref{f3} (solid line) shows a typical
example: a steep gaussian-like increase, a pronounced maximum, and a
slow exponential-like decrease. There seems to be no universal
distribution function which can be rescaled for different values of
$N$, $b$, and $g$ in order to fit the data. Therefore no scaling
laws for the width of the distribution are expected. We only found
that the increase is stronger than $N^2$ but weaker than $N^3$.
Hence the relative width decreases with $N$. The dotted
line in Fig.~\ref{f3} shows the energy distribution of the states
into which stable states having very large energies relax after the
first avalanche. The distribution is a sharp peak at the position
near the steep gaussian-like increase of the distribution of the
energy of the SOC attractor. This fact is consistent with the
observation that very large avalanches of states of the attractor with
energies large above the mean value leave the system in a state almost at
the place of the sharp peak (see Fig.~\ref{f3}).

A typical distribution of $x_i\,\mbox{mod}\,1$ is shown in
Fig.~\ref{f6}. An enlargement shows the self-similar character of
this distribution.  If we would generate stationary states by
iterating the map (\ref{map1}), all positions would be equally
probable because the invariant density is uniform (see
Sec.~\ref{s2}). The driving mechanism selects states with a potential
energy from a relatively small subinterval. Thus the selecting
mechanism forbids certain symbols or symbol sequences. For example,
for $b=1$ and $g=0.1$ the $f$ symbols $\pm 2$ become almost forbidden
symbols [the probabilities are $p(-2)=2.1\cdot 10^{-4}$ and
$p(2)=4.7\cdot 10^{-3}$] whereas in the case of no selection $p(\pm
2)=1/8$. The loss of symbols is reflected by the loss of possible
values for $x_i\,\mbox{mod}\,1$ because of (\ref{ssl}). The
fractality of the distribution is a well-known consequence of this
fact (note, e.g., the classical construction of the Cantor set, where
one of three symbols is forbidden). Note that $x_i\,\mbox{mod}\,1$ is
not uniquely given by the $f$-symbol sequence; we need
$x_i\,\mbox{mod}\,1$ and $x_{i+1}\,\mbox{mod}\,1$ in order to get the
$f$-symbol sequence. Thus Fig.~\ref{f6} is in fact a projection of a
two-dimensional fractal distribution.

\section{Avalanche statistics and the dynamics on the SOC attractor
}\label{s3}

After presenting the statistical properties of the SOC attractor we
investigate the reduced dynamics on it. First, we are looking at
the distribution of avalanches.

There are several properties of an avalanche which can be measured:
\begin{description}
\item[Strength:] There are two ways to measure the strength of an
avalanche:
\begin{itemize}
\item The most natural way is to measure the difference $\Delta E_{pot}$
of the potential energies just before and after the avalanche. Note
again that for most avalanches (i.e. for all small
avalanches) $\Delta E_{pot}$ is less than the energy put into the
system between two successive avalanches (see Fig.~\ref{f3}).
\item Another way usually used in connection with the BK model is to
sum over the displacements of all particles before and after the
avalanche
\begin{equation}
  S=\sum_{i=1}^N |x_i^{after}-x_i^{before}|.
  \label{S}
\end{equation}
If $x_i^{after}>x_i^{before}$ holds for all $i$ then $S/N$ is
equivalent to the displacement of the center of mass.
\end{itemize}
\item[Length:] Almost all avalanches start at the first
particle (only for large values of $g$ there is a considerable amount
of avalanches starting at the second or third particle). Therefore we
can define the length $L$ of an avalanche as the index of the
furthest particle changing its well number:
\begin{equation}
  L=\max\{i\in\{1,\ldots,N\}|n_i^{after}\neq n_i^{before}\}
  \label{L}
\end{equation}
\item[Duration:] As the duration $D$ of an avalanche we define the
time needed to fulfill the stopping criterion (see Sec.~\ref{s2a}).
\end{description}
In the following we present cumulative densities of all these
properties. The cumulative density $p(x)$ is the probability $p$ to
find an event $y$ greater than $x$. If it is a power law of the form
$p(x)\propto x^{-B}$ the event distribution is also a power law, i.e.
$\rho(x)\equiv dp/dx\propto x^{-B-1}$.

Figs.~\ref{f7}a-c show cumulative densities of avalanche strength for
$b=1$, $b=5$, and $b=20$, respectively. All curves are shifted by
$\log_{10}g$. We do not find pure
power laws. There are of course finite-size cutoffs for very strong
avalanches and steps for very weak avalanches caused by the
discreteness of the reduced dynamics. But the intermediate parts
do not show nice straight lines. For low damping values we see a
pronounced cross-over on Figs.~\ref{f7}a and \ref{f7}b. Above the
cross-over (i.e. for strong avalanches) a power law emerges which
becomes visible only if the number of particles $N$ is large enough.
In Fig.~\ref{f7}a, for example, it is not seen for $N=125$ because $N$
is too small. Presumably the power law found for $N=500$
continues for larger $N$ and the finite-size cutoff shifts
to higher values. This is confirmed by a simulation with 2000
particles for $b=1$ and $g=0.3$.
The exponents $B^{(\Delta E_{pot})}$ and $B^{(S)}$ varies between 0.4 and
0.6. Similar values are found for the FK models treated in \cite{r7,r8}.
For large values of $b$ and small damping values $B^{(\Delta E_{pot})}$
becomes larger than $B^{(S)}$. An example is shown in Fig.~\ref{f7}c for
$b=20$ and $g=0.01$. Here $B^{(\Delta E_{pot})}\approx 0.8$ and
$B^{(S)}\approx 0.5$.

Figs.~\ref{f8}a-c show cumulative densities of avalanche length and
duration. Distributions of length are qualitatively similar to
distributions of strength. There is also a cross-over for $b\leq 5$
and small damping. The cross-over shifts to lower values of $L$ if
$b$ increases. Above the cross-over (i.e. for large avalanches) the
power-law exponent $B^{(L)}$ is roughly equal to unity. The
significant drop at $L=N$ means that the probability of an avalanche
affecting the whole chain $p(L=N)$ is much larger (usually by two
orders of magnitude) than the probability for the largest avalanche
which just stops before the end. The absolute value of $p(L=N)$
varies between $2\cdot 10^{-3}$ and $2\cdot 10^{-2}$. Distributions
of duration show power laws only for $b=1$, $g=0.1$ and $g=0.3$ with
$B^{(D)}\approx 1.5$ and $B^{(D)}\approx 1.3$, respectively.

In order to study the dynamics on the attractor it would be useful to
have some visualization of the trajectories on it. One way is to
print a list of $f$-symbol sequences representing successive states.
But it would be difficult to interpret this list. For
example, a jump of one particle into its neighboring potential well
not only leads to a change of its own $f$ symbol but also of the
$f$ symbol of its neighbors. Therefore we choose a different
representation. From Sec.~\ref{s4} we know that the position of each
particle is roughly given by $x_i\approx x_0-a\,i+i(i-2N-1)\overline{f}/2$.
If this relation is exact, $x_i+a\,i+i(2N+1-i)\overline{f}/2$ plotted
as a function of $i$ would give straight lines. Fig.~\ref{f9} shows what
actually happens: we get wiggly lines instead of straight lines.
Since an avalanche usually involves only the part of the chain near its
fixed end, these lines build up a tree-like structure. The height of
a branching point is the length $L$ of the avalanche whereas the area
between two lines is the strength $S$. An inspection of
the figures shows that almost always $x_i(\tau+1)>x_i(\tau)$.
Therefore $S/N$ is the displacement of the center of mass.

{}From Fig.~\ref{f9} we also see that, on average, $x_0$ has to be
shifted by a finite amount in order to trigger the next avalanche.
Because of (\ref{x0}) $x_0$ is proportional to the time. Therefore
$x_0(\tau+1)-x_0(\tau)$ is the waiting time between successive
avalanches measured in time units given by the inverse pulling speed
$v$. Fig.~\ref{f10} shows distributions of $x_0(\tau+1)-x_0(\tau)$.
They are also self-similar like the
distributions of $x_i\,\mbox{mod}\,1$ (see Fig.~\ref{f6}). From
Sec.~\ref{s2} we know that $x_0(\tau+1)-x_0(\tau)<x_0^{right}-x_0^{left}=
1/\eta$. Choosing stationary states randomly, $x_0$
would be equally distributed between $x_0^{left}$ and $x_0^{right}$.
Therefore we expect the mean value $\overline{x_0(\tau+1)-x_0(\tau)}=
\eta^{-1}/2$. From the simulation
where a specific selection takes place we found that the mean value
is always less than $\eta^{-1}/2$. It increases with decreasing
damping and seems to approach this value for $g\to 0$.

One might expect that a tiny shift of $x_0$ may trigger an avalanche.
But this not the case (see from Fig.~\ref{f10}). There is
a gap near zero where the probability to find a value for
$x_0(\tau+1)-x_0(\tau)$ in this gap is almost zero.
Although the system is driven far away from equilibrium
and a huge amount of energy is stored in the chain, a {\em finite\/}
amount of energy is needed in order to trigger an
avalanche. For large values of the damping constant the gap shrinks
to zero and a finite density at zero appears. As a consequence we
also find avalanches which start not at the first particle but at
particles deeper in the chain.

{}From experiments on two objects which are in dry contact and which are
moved very slowly against each other it is well known that the pulling force
fluctuates. It is expected that the power spectra of these
fluctuations show a power law $1/f^{B^{(F)}}$ \cite{r22b,r23}.
Fig.~\ref{f11} shows the pulling force as a function of $x_0$. The
pulling force is $x_0-x_1-a$. The jumps correspond to avalanches. The
power law shows clearly $1/f$ noise. For weaker damping the exponent
$B^{(F)}$ increases and seems to reach 2 for $g\to 0$. The oscillations in the
power spectra are caused by the nearly periodic appearance of the
avalanches. Therefore the frequency of the first peak is roughly given by
$1/\overline{x_0(\tau+1)-x_0(\tau)}$. These results are independent of
the chain length $N$ and nearly independent of the potential strength
$b$.

\section{Details of single avalanche dynamics}\label{s6}

In this section we take a closer look at the dynamical process of a
single avalanche. This will give us some insight into the
question how the chain is able to organize itself into a critical
state.

Figs.~\ref{f12} and \ref{f13} shows for $b=1$ and $b=5$, respectively, two
typical examples of large avalanches which do not hit the end of the
chain. The dynamics are visualized by (a) kinetic energy plots where a
grey scale shows the kinetic energy of each particle averaged over a
time period given by the smallest phonon frequency, and by (b)
displacement plots where for equidistant time steps the displacement
of each particle from its initial value is drawn.
We easily see two type of propagations:
\begin{description}
\item[Sound waves:] This are the smooth waves at low level in the
kinetic energy plots. They are {\em linear\/} waves.
{}From the phonon dispersion relation (\ref{wk}) we
immediately calculate the sound velocity $v_S$ as
\begin{equation}
  v_S=\max_k\frac{d\omega_k}{dk}=\eta^{1/2},
  \label{svel}
\end{equation}
where $\eta$ is given by (\ref{eta}). This formula agrees very well with
the sound waves found in the simulations.
\item[Shock waves:] This are the sharp waves at high level in the
kinetic energy plots. They are {\em nonlinear\/} waves of propagating
hopping events. This is easily seen in the displacement plots of the
first shock wave which starts at the first particle and propagates to
the right. Shock waves can stop at certain points. They can be
partially or totally reflected at some points. Thus they propagate in
different direction and most particles will be hit by more than one
shock wave. A shock wave is often accompanied by several delayed
shock waves propagating in the same direction.  The velocity is
not constant during a shock wave but varies only slightly. However
it is always larger than the sound velocity. The driving mechanism
is the release of energy during particle hopping.  We give a rough
estimate of the released energy $E_{rel}$ of a single particle
hopping: compare two configurations where only one particle changes
its position from $x_i^{before}$ to $x_i^{after}$. Neglecting the
energy of the external potential we get $E_{rel}=(x_i^{after}-x_i^{before})
(x_{i+1}+x_{i-1}-x_i^{after}-x_i^{before})$. On average, we get
\begin{equation}
  E_{rel}\approx \overline{d}\cdot\overline{f},
  \label{Erel}
\end{equation}
where $\overline{d}$ is the averaged displacement $x_i^{after}-x_i^{before}$
and $\overline{f}$ is the mean local stress of the chain introduced in
Sec.~\ref{s4}. Shock waves are possible only if the local stress of the
chain is unequal zero.
\end{description}

It is expected that the SOC attractor is a critical state in the
sense that, on average, perturbations (i.e.  avalanches) are {\em
just\/} able to propagate through the whole system \cite{r5}. In the
sand-pile metaphor the slope of the pile is the dynamical ``control
parameter'' which moves into the critical point. The corresponding,
self-organizing ``control parameter'' in our model is the {\em
Peierls barrier\/} of the shock waves which they feel because of the
discreteness of the chain. The barrier height is proportional to the
strength of the external potential $b$, it decreases with increasing
local stress $\overline{f}$, and it shrinks to zero for $\overline{f}\to
b$ which corresponds to the stable stationary state with the largest
energy. It is difficult to quantify the barrier height. Instead, we
use the mean local stress $\overline{f}$.

In order to understand why the Peierls barrier is the self-organized
``control parameter'' we introduce a much simpler model, where the
analogon is a control parameter. The model is a single damped
particle on a tilted ``washboard'': the slope $f$ of the washboard is
the control parameter. There is also a barrier which decreases with
increasing slope. Thus, the slope $f$ corresponds to the
mean local stress $\overline{f}$. Propagating particle means shock
wave and pinned particle means stationary state. The key property of
the washboard model and FK model is the {\em bistability\/} between
propagation and pinning due to {\em inertia\/}. Bistability is
possible only if the control parameter is above a critical value
$f_c$. The question is now: does the weakly driven FK model organize
its ``control parameter'' (i.e. the mean local stress) in such a way
that $\overline{f}\approx f_c$, where an avalanche is just able to
propagate through the whole chain?

In order to answer this question numerical experiments of the
following type were done: a chain with 500 particles is prepared in a
state with a periodic $f$-symbol sequence. The beginning of the chain
(i.e. the first 15 particles) is ``overloaded'' which means that the
mean local stress is much larger than for the rest of the chain. For
the same initial state but varying damping constant $g$ it is
determined whether the avalanche propagates until the end of the
chain or not. The experiments were done for $b=1$. The $f$-symbol
sequences were generated by a Markov process with transition
probabilities derived from the distribution of two-symbol words taken
from actual simulations for $g=0.03$ and $g=0.3$. The lengths of the
$f$-symbol periods were $50$ and $40$, respectively. The experiments
give clear evidence for the existence of a $g_c$ with the property
that for $g>g_c$ all avalanches are always stop before reaching the
end of the chain, and for $g<g_c$ they always reach the end. The
filled squares and triangles in Fig.~\ref{f14} show the relation
between $g_c$ and the mean local stress $\overline{f}$. The small
fluctuations are caused by the fact that different $f$-symbol
sequences with the same $\overline{f}$ may have slightly different
Peierls barriers. Nevertheless we can draw a curve $f_c(g)$ through
these points. Above this curve shock waves propagate. The open and
larger squares in the figure denote the mean local stress from the
SOC simulations.  They lie directly on the curve.  Therefore the FK
model selects a mean local stress for which a perturbation is just
able to propagate through the whole chain.

\section{The transition to regular behavior}\label{s5}

For large $g$ the SOC behavior will eventually disappear and regular
behavior takes place. It is characterized by periodic orbits where
always the same states periodically reappear only shifted by some
integer amounts of the periodicity of the external potential. The
simplest orbit has period one, and the avalanche is a single
shock wave propagating from the pulled end to the free end of the
chain. Orbits with higher periods contain at least one avalanche with
$L=N$. Typically a large avalanche is accompanied by very small
ones ($L<10$).

There is no sharp transition from regular to irregular behavior. For
example for $b=1$ and $N=125$ period-two orbits were found for
$g=0.28$ and $g=0.4$, whereas for $g=0.9$ in a particular simulation
no periodicity was found after $2\cdot 10^4$ avalanches. Thus, there
is a broad transition regime ($g\approx 0.28\ldots 0.95$ for $b=1$)
of coexistence of regular and irregular behavior. There also is
multistability between different periodic orbits. The number of
periodic orbits and their basin of attraction (in the reduced phase
space) seems to increase with $g$. It is difficult to decide whether
an irregular orbit is indeed aperiodic (i.e. having a period
comparable to the recurrence time) or whether it is only a transient. The
avalanche statistics of the irregular cases deviate from the behavior
shown in Sec.~\ref{s3}. The degree of deviation increases with $g$.
For $g=0.9$ the distribution of the large avalanches show an
exponential law instead of a power law. One reason may be an
{\em intermittent\/} behavior where the erratic sequence of
avalanches is interrupted by a regular episode of the following type:
similar to the periodic case, one large shock wave travels through the
chain up to a certain point and shifts all particles by the same
integer amount. The stopping point comes from avalanche to avalanche
nearer and nearer to the beginning of the chain. The regular episodes
last at most until the stopping point reaches the beginning of the chain.
Again these large avalanches may be accompanied by tiny ones.

In order to understand the behavior in this transition regime we look
again at the single avalanche dynamics. Fig.~\ref{f15} shows an
example for $g=0.9$. The sound waves are nearly suppressed, and only shock
waves are visible. At certain points they stop, turn their direction
of propagation or split into two waves. The $f$-symbol sequence of the
initial state of this particular example is the periodic pattern
$\{0,1,0,1,\cdots\}$ with some imperfections. For example around
$i=35$ it is changed into $\{\cdots,0,1,0,1,1,0,1,1,0,0,1,0,1,\cdots\}$.
At that imperfection the initial shock wave splits into two waves (see
Fig.~\ref{f15}a). Before it reaches the splitting point it shifts all
particle by unity (see Fig.~\ref{f15}b). A similar shift happens during
the propagation of the backward travelling wave.

We conclude from the inspection of Fig.~\ref{f15}b that almost all
shock waves are {\em regular\/}, although the overall dynamics may be
irregular. There are two types of regular shock waves: the first type
turns a regular state (represented by a periodic $f$-symbol sequence)
into the same state with the same $f$-symbol sequence only shifted by
an integer. The other type turns a regular state into another regular
one with a different $f$-symbol sequence but the same mean
local stress. Regular shock waves are periodic solutions of the equation
of motions (\ref{eqm}) with given boundaries for $i\to\pm\infty$. For
example, a shock wave of the first type which shifts the state
$\{\cdots,1,1,1,\cdots\}$ by unity is given by
$x_i(t)=n_i+x^{(0)}+\xi(i-c\,t)$, where $x^{(0)}=(1-1/b)/2$, $c$ is
the velocity, and $\xi$ is a solution of the delay-differential
equation
\begin{eqnarray}
 c^2\ddot{\xi}(t)-c\,g\dot{\xi}(t)&=&\xi(t+1)+\xi(t-1)-2(1+b)\xi(t)
  \nonumber\\ &&+2b\,\Theta(\xi(t)-1+x^{(0)}),
  \label{dde}
\end{eqnarray}
where $\Theta$ is Heaviside's step function. The boundary conditions are
$\xi(t\to\pm\infty)=(1\pm 1)/2$. Such a regular shock wave is
possible only if (\ref{dde}) has a solution. Any regular shock wave
is a solution of a similar boundary value problem of a
delay-differential equation (or a set of them). If $g$ is larger than
a critical value $g_c$ the boundary value problem will only have a
solution for $c=0$, i.e. a stationary kink-like structure.
This is the same $g_c$ as in Sec.~\ref{s6}.

Instead of solving such boundary value problems, numerical studies
were performed by preparing both halves of the chain into the states
defined by the boundary conditions. In the case of regular shock
waves, the interface in the middle of the chain evolves into a single,
confined shock wave if $g$ is just below $g_c$. The state behind the
shock wave may not be the same as given by the boundary values if the
shock wave is of type two. We found a second critical value
$g_c^r<g_c$ below which the regular shock wave turns into an
irregular one. A irregular shock wave is distinguished from a regular
one by the fact that the particles behind it do not relax into a
regular state. Instead they move further than it is necessary to build up
a state with same mean local stress as before the shock wave. This
motion caused by inertia tends to decrease the mean local stress.
Usually additional shock waves will be generated, some
propagating backward, some forward but delayed. The state behind
the initial shock waves become unpredictable although the initial
state is regular. From the numerical experiments it is not possible
to decide whether the irregular shock waves are caused by an
instability or by the disappearance of the regular ones. Almost all
shock waves in Figs.~\ref{f12} and \ref{f13} are of the irregular
type.

{}From Sec.~\ref{s6} we already know that $g_c$ is a monotonically
increasing function of the mean local stress $\overline{f}$. The same
property holds for $g_c^r$. Numerical observations lead to the
conjecture that regular shock waves are not possible (i.e.
$g_c^r=g_c$) for mean local stresses below a critical value
$\overline{f}_c$. The corresponding value of $g_c$ where this
happens defines the lower bound of the transition regime $g_T^{low}$.
Thus for $g<g_T^{low}$ the inherent chaotic behavior of the irregular
shock waves prevents the dynamics in the reduced phase space to
become regular. The irregular behavior for $g>g_T^{low}$ are either
caused by irregular shock waves or by irregularities of the initial
state. These irregularities are unavoidable even for regular initial
states because they are generated by the regular shock waves at
boundaries. But a careful construction
of the initial states makes it possible to find small periodic orbits
in the reduced phase space even for very low values of $g$. The above
mentioned orbits for $g=0.28$ and $g=0.4$ were constructed in this
way. The $f$-symbol sequences are
$\{(0,0,1)^2,(0,0,0,1,0,0,1)^{17}\}$ and $\{(0,0,1)^{41},0,1\}$,
respectively \cite{foo2}. The other parameters are $a=1.3$, $b=1$,
and $N=125$. These orbits will be stable if the parameters of the
systems are changed only slightly. But they are unstable if
only one symbol is changed.

There seems to be no clear definition for the upper bound of the
transition region. For increasing $g$ the probability for irregular
shock waves is reducing eventually to zero. But as
mentioned above this does not exclude irregular behavior. The
following observation may be important: for $b=1$ and $g=0.95$ a huge
number of states (but not all!) of the form
$\{(1,0)^{m_1},1,(1,0)^{m_2},1,\ldots,(1,0)^{m_M},1,1\}$ are
reproduced by an orbit of period 1 or 2. Thus, the upper bound may be
defined by the point where an {\em irregular\/} state is reproduced
by a {\em regular\/} shock wave.

\section{Conclusion}\label{scon}

In this paper the self-organization due to weak local driving (i.e.
slow pulling of one end of the chain) of the damped FK
model with a piecewise parabolic potential was studied in detail. The
piecewise parabolic potential makes it possible to drive the system
infinitesimally slowly because stability intervals of all stationary
states are computable. Therefore, as in ``sand pile'' models, a
separation of the dynamics between driving (putting the system at the
edge of the instability) and relaxing due to an avalanche is
possible.

The driving mechanism leads to an attractor which is characterized by
the fact that, on average, the difference between successive springs
(i.e. $x_{i+1}-2x_i+x_{i-1}$) called local stress is equal for each
particle in the bulk of the chain. The mean local stress corresponds
to the mean slope of a sand pile. And, as in those ``sand pile''
models, it also organizes itself into a critical state where a
perturbation is just able to propagate through the whole chain. The
mean local stress is the self-tuned ``control parameter'' of the
system, because particle hopping releases energy by reducing local
stress. The released energy drives shock waves of particle-hopping
events which build up avalanches. The mean local stress of the chain
has the same effect as a constant externally controlled force applied
to each particle. The coexistence between propagation and pinning due
to inertia is responsible for SOC.

The chain must be relatively long ($N>125$) in order to get power
laws of the distributions of avalanche strengths and lengths.  The
weaker the potential the longer it must be. The discreteness of the
model means that SOC is a statistical property clearly visible only
in the thermodynamic limit. We have measured the cumulative
distributions of the avalanche strength calculated either by the
energy drop $\Delta E_{pot}$ or by the displacement sum $S$ (in
almost all cases $S/N$ is the displacement of the center of mass). In
both cases the exponent is roughly 1/2, whereas it is roughly 1 for
the avalanche length. We have also measured the power spectra of the
fluctuating pulling force. The exponent increases from 1 to 2 with
decreasing damping constant.

If dissipation is strong enough irregular behavior of the SOC regime
will turn into a regular one. In the regular regime a sequence of
states reappears periodically only rigidly shifted by an integer
multiple of the periodicity of the external potential. The transition
between irregular and regular behavior is relatively broad. It is
characterized by the coexistence between irregular behavior and
several periodic orbits. In the irregular case the type of avalanche
distributions is smoothly changing from a power law at the lower edge
of the transition regime to an exponential law at the upper edge.
The irregular non-SOC behavior is accompanied by a kind of
intermittency.

As mentioned in the introduction, the FK model is very similar to the
BK model and in fact it becomes the BK model for $b\to\infty$. In
\cite{r11} the BK model was also driven by slow pulling of one end of
the chain (train model). The exponents of the power laws of the
avalanche distributions are in good agreement with our results from
the FK model. Preliminary simulations have shown that we
also get similar results as in
\cite{r9a,r9b} if the FK model is driven like the BK model in those papers
(i.e. all particle are connected with a slowly
moving rigid plate via soft springs). Especially the same kind of phase
transition as in \cite{r9b} occurs if the friction parameter is
varied.
In \cite{r9b} it is argued that in such uniformly driven BK models
criticality needs some fine tuning rather than it is self-organized.

In \cite{r20} the FK model was also driven by slowly pulling one end
of the chain. But the external potential was given by randomly
distributed pinning centers, and the dynamics was a kind of overdamped
dynamics. That is, instead of simulating the avalanche dynamics the
next state was chosen by the rule to take the state with the smallest shift
of the center of mass. The distributions of avalanche strength
measured by the energy drop gives a power law with an exponent of
roughly 0.8, i.e. the half of the value we obtained. The distributions of
the center-of-mass displacements do not show any power law. The power
spectra of the pulling force shows an $1/f^{1.5}$ law which is roughly
similar to our results.

Concerning the conditions for SOC there remain a number of of open
questions: (i) What is the role of a random external potential? (ii)
What are the effects of different types of driving (uniform versus
nonuniform, reducing the number of meta-stable states
versus leaving it constant)? (iii) For which kind of models is inertia
important? (iv) What is the role of the dimensionality of the model?

Another important and open question not concerning SOC is: what
happens for finite pulling velocities? For small velocities
avalanching will be still the dominant behavior as long as the averaged
waiting time between successive avalanches is smaller than the averaged
avalanche duration. For larger velocities the chain will never be at
rest. For small values of the damping constant phonon resonances are
expected \cite{r24}. For very large velocities a strong damping force
will dominate the force from the external potential. Thus we expect
the same behavior mentioned in Sec.~\ref{s4}, paragraph~7.

\acknowledgments
I gratefully acknowledge helpful discussions with T. Christen, S. Fauve,
J. Langer, R. Schilling, M. de Sousa Vieira, T. Strunz, H. Suhl, P.
Talkner and H. Thomas. I also acknowledge the possibility to do the
simulations on the NEC SX-3 at the Centro Svizzero di Calcolo
Scientifico at Manno, Switzerland.
This work was supported by the Swiss National Science Foundation.

\begin{figure}
\caption{The Frenkel-Kontorova (FK) model with piecewise parabolic
external potential $V_E$. An example of a stationary state and its well
numbers and $f$ symbols are shown. The state is stable for adiabatic
moves of the fixed end of the chain (i.e. the zeroth particle)
between $x_0^{left}$ and $x_0^{right}$.}
\label{f1}
\end{figure}

\begin{figure}
\caption{The fixed end of the chain $x_0$ as a function the free end
$x_N$. The parameters are $a=1.3$, $b=1$, and $N=4$.}
\label{f2}
\end{figure}

\begin{figure}
\caption[f3]{Transient and SOC attractor of the reduced dynamics. The
parameters are $b=1$, $N=125$, and $g=0.1$. The initial
state is defined by $f_i=0$ for $i=1,\ldots,N$. The first avalanche
which involves the whole chain is marked by an arrow. The energy
distributions of the SOC attractor and of 5500 relaxation experiments
with $\overline{f}=0.6$ (for details, see text) are denoted by a solid
line and a dotted line, respectively.
}
\label{f3}
\end{figure}

\begin{figure}
\caption{Temporal average of $x_i\,\mbox{mod}\,1$ over 15000 avalanches.
The parameters are $b=1$, $N=125$, and $g=0.1$. The dotted
line is the expected value assuming no $i$ dependence and using
Eqs.~(\protect\ref{en3}) and (\protect\ref{lx}).}
\label{f4}
\end{figure}

\begin{figure}
\caption{The mean potential energy $\overline{E}_{pot}$ as a
function of the damping constant $g$. The results are obtained from
simulations with $N=500$ and $N=125$ for $b\leq 5$ and $b=20$,
respectively.}
\label{f5}
\end{figure}

\begin{figure}
\caption[f6]{Distribution of $x_i\,\mbox{mod}\,1$ averaged over the
bulk particles (i.e. from $i=10$ until $i=115$). Parameters are
$b=1$, $N=125$, and $g=0.1$. The larger peaks are marked by sequences
of sums of $f$ symbols. A sequence is defined by
$f_i,f_{i-1}+f_{i+1},f_{i-2}+f_{i+2},\ldots$. The square bracket
denotes the width of possible values assuming $f_j\in\{-1,0,1\}$.}
\label{f6}
\end{figure}

\begin{figure}
\caption{Cumulative densities of avalanche strengths.
The curves are denoted by $g$ and they are shifted by $\log_{10}g$.}
\label{f7}
\end{figure}

\begin{figure}
\caption{Cumulative densities of avalanche lengths and durations for
(a,b) $N=500$ and (c) $N=125$.
The curves are denoted by $g$ and they are shifted by $\log_{10}g$.}
\label{f8}
\end{figure}

\begin{figure}
\caption{Examples of trajectories on the SOC attractor for (a) $b=1$, $N=125$,
and $g=0.1$ and (b) $b=5$, $N=125$, and $g=0.03$. The states
just before an avalanche are drawn. The mean shape of the chain
$i(i-2N-1)\overline{f}/2-a\,i$ is subtracted from $x_i$, where (a)
$\overline{f}=0.138$ and (b) $\overline{f}=0.187$.}
\label{f9}
\end{figure}

\begin{figure}
\caption{Distribution of $x_0(\tau+1)-x_0(\tau)$ for (a) $b=1$,
$N=250$, and $g=0.1$ and (b) $b=5$, $N=125$, and $g=0.1$. It is also the
distribution of the waiting time between successive avalanches in accordance
with Eq.~(\protect\ref{x0}). The arrow denotes the mean value.}
\label{f10}
\end{figure}

\begin{figure}
\caption{Pulling force and its power spectrum for (a) $b=1$, $N=250$, and
$g=0.1$ and (b) $b=5$, $N=125$, and $g=0.1$. The line has the slope -1.1
and -1.2 for (a) and (b), respectively. The arrows denote the frequencies
corresponding to the mean distance between two successive jumps
of the pulling force. }
\label{f11}
\end{figure}

\begin{figure}
\caption[f12]{
A time-resolved single avalanche for $b=1$, $N=500$, and $g=0.03$.
Avalanche parameters: $\Delta E_{pot}=2.65\cdot 10^3$, $S=4.37\cdot 10^4$,
$L=386$, and $D=1.05\cdot 10^3$.
(a) Density plot of the kinetic energy averaged over $2\pi/\sqrt{2b}$.
(b) Displacement curves $x_i(t)-x_i(0)$ for equidistant time steps $\Delta
t$ with $\Delta t=20$.}
\label{f12}
\end{figure}

\begin{figure}
\caption[f13]{
A time-resolved single avalanche for $b=5$, $N=500$, and $g=0.03$.
Avalanche parameters: $\Delta E_{pot}=1.67\cdot 10^4$,
$S=8.96\cdot 10^4$, $L=435$, and $D=9.45\cdot 10^2$.
(a) Density plot of the kinetic energy averaged over $2\pi/\sqrt{2b}$.
(b) Displacement curves $x_i(t)-x_i(0)$ for equidistant time steps $\Delta
t$ with $\Delta t=20$.}
\label{f13}
\end{figure}

\begin{figure}
\caption[f14]{Bistability regime of shock waves (filled symbols) and
self-organized mean local stress (open squares) for $b=1$. Filled
squares and triangles denote the bistability edge for different
states with periodic $f$-symbol sequences generated from two-symbol
distributions taken from simulations with $g=0.3$ and $g=0.03$,
respectively (for more details, see text).}
\label{f14}
\end{figure}

\begin{figure}
\caption[f15]{
A time-resolved single avalanche for $b=1$, $N=125$, and $g=0.9$.
Avalanche parameters: $\Delta E_{pot}=7.89\cdot 10^2$,
$S=1.47\cdot 10^3$, $L=113$, and $D=4.19\cdot 10^2$.
(a) Density plot of the kinetic energy averaged over $2\pi/\sqrt{2b}$.
(b) Displacement curves $x_i(t)-x_i(0)$ for equidistant time steps $\Delta
t$ with $\Delta t=5$.}
\label{f15}
\end{figure}


\begin{references}
\bibitem{foo1}
  If the cusps of the external potential are slightly rounded the
  jumps would be replaced by curves of unstable states connecting the pieces
  of stable states. Thus the end points of the stable pieces are
  related to saddle-node bifurcations.
\end{references}
\end{document}